\documentstyle[12pt,epsfig]{article}
\begin{document}
\noindent
\begin{center}
{\Large {\bf Coincidence Problem
and Holographic \\$f(R)$ Gravity in Spatially flat and Curved Universes}}\\ \vspace{2cm}
 ${\bf Yousef~Bisabr}$\footnote{e-mail:~y-bisabr@srttu.edu.}\\
\vspace{.5cm} {\small{Department of Physics, Shahid Rajaee Teacher
Training University,
Lavizan, Tehran 16788, Iran}}\\
\end{center}
\vspace{1cm}
\begin{abstract}
The $f(R)$ gravity models formulated in Einstein conformal frame
are equivalent to Einstein gravity together with a minimally
coupled scalar field. The scalar field couples with the matter sector and the coupling term is given by the conformal factor.
We apply the holographic principle to such an interacting model in spatially flat and curved universes.  We show that the model leads to a constant ratio of energy densities of dark matter to dark energy in a spatially flat universe.  In a spatially curved universe, the ratio is not a constant
and the evolution seems to be model-dependent.  However, we argue that any cosmologically viable $f(R)$ model can lead to a nearly constant ratio of energy densities and therefore alleviate the coincidence problem.

\end{abstract}
\vspace{3cm}
It is now generally believed that the universe is currently undergoing a period of accelerated expansion.
The simplest candidate to produce this cosmic
acceleration is the cosmological constant, the energy density
associated with quantum vacuum.  However, there are several problems for associating cosmic acceleration with the cosmological
constant.  First, theoretical estimates on its value are many
order of magnitude larger than observations \cite{win}.  Second,
it is simply a constant, namely that it is not diluted with
expansion of the universe.  This latter is specifically important
in the sense that there are observational evidence \cite{R}
demonstrating that the cosmic acceleration is a recent phenomena and
the universe must have passed through a deceleration phase in the
early stages of its evolution.  This deceleration phase is
important for successful nucleosynthesis as well as for the
structure formation. We therefore need a field evolving during
expansion of the universe in such a way that its dynamics makes
the deceleration parameter have a signature flip from  positive in
the early stages of matter dominated era to negative in the
present
stage \cite{b}.  There is also another
problem which is the focus of the present note.  It concerns with the coincidence between the observed vacuum
energy density and the current matter density.  While these two
energy components evolve differently as the universe expands,
their contributions to total energy density of the universe in the
present epoch are the same order of magnitude.  \\
As a different point of view, cosmic acceleration may be interpreted as evidence either for
existence of some exotic matter components or for modification of
the gravitational theory.  In the first route of interpretation
one can take a mysterious cosmic fluid with
sufficiently large and negative pressure, dubbed dark energy.
These models are usually invoked a scalar field which during its
evolution takes negative pressure by rolling down a proper
potential.  In the second route, however, one attributes the
accelerating expansion to a modification of general relativity. A
particular class of models that has recently drawn a significant
amount of attention is the so-called $f(R)$ gravity models
\cite{r}. These models propose a modification of
Einstein-Hilbert action so that the scalar curvature is replaced
by some arbitrary function $f(R)$. \\
Recently, different models inspired by holographic principle have been proposed to explain the cosmic acceleration.
The basic
idea is that the number of degrees
of freedom of a physical system scales with its bounding area
rather than with its volume \cite{suss}.  For an effective quantum field
theory in a box of size $L$ with an ultraviolet (UV) cutoff
$\Lambda$, the entropy $S$ scales extensively as $S\sim
L^{3}\Lambda^{3}$. However the peculiar thermodynamics of black
holes has led Bekenstein \cite{bek} to postulate that the maximum
entropy in a box of volume $L^{3}$ behaves non-extensively,
growing as the area of the box.  In this sense there is a
so-called Bekenstein entropy bound
\begin{equation}
S=L^{3}\Lambda^{3}\leq S_{BH}\equiv \pi L^{2}M^{2}_{p}
\label{0a}\end{equation} where $S_{BH}$ is the entropy of a black
hole of radius $L$, and $M_{p}\equiv (8\pi G)^{-\frac{1}{2}}$
stands for the reduced Planck mass.  It is important that in this
relation the length scale $L$ providing an Infrared (IR) cutoff is
determined by the UV cutoff $\Lambda$ and can not be chosen
independently. However such a non-extensive scaling law seems to
provide a breakdown of quantum field theory at large scales.  To
reconcile this breakdown with the success of local quantum field
theory in describing observed particle phenomenology, Cohen et al.
\cite{co} proposed a more restrictive bound.  Since the maximal
energy density in the effective theory is of the order
$\rho_{\Lambda}=\Lambda^{4}$, requiring that the energy in a given
volume not to exceed the energy of a black hole of the same size
results in the constraint
\begin{equation}
L^{3}\rho_{\Lambda}\leq L M^{2}_{p} \label{02}\end{equation} If we
take the largest value of the length scale $L$ as the IR cutoff
saturating the inequality (\ref{02}), we then obtain the
holographic dark energy density
\begin{equation}
\rho_{\Lambda}=3c^{2} M^{2}_{p}L^{-2} \label{03}\end{equation} in
which $3c^{2}$ is a numerical constant.  It is interesting to note
that if the length scale $L$ is characterized by the size of the
universe, the Hubble scale $H^{-1}$, then equation (\ref{03})
gives a vacuum energy density of the right order of magnitude
consistent with observations \cite{co}.  It is however pointed out
that this yields a wrong equation of state parameter for dark
energy, and other possible values for $L$ should be chosen such as
the size of the future event horizon \cite{li}.  This conclusion
is, however,  based on the assumption of an independent evolution
of energy densities of dark energy and dark matter.  It is shown
\cite{pp} that, if there is \emph{any} interaction between these two
components the identification of $L$ with $H^{-1}$ is possible.  In particular, the authors of \cite{pp} \cite{ref} argued that such an
identification
necessarily implies a constant ratio of the energy densities of the two components \emph{regardless of the details of the interaction}. \\
In the present note, we investigate the coincidence problem in the context of holographic  $f(R)$ gravity models.
In $f(R)$ models the dynamical variable of the vacuum sector
is the metric tensor and the corresponding field equations are fourth order.  This
dynamical variable can be replaced
by a new pair which consists of a conformally rescaled metric and a scalar partner.  Moreover, in terms of the new set
of variables the field equations are those of General Relativity.  The original set of variables
is commonly called Jordan conformal frame and the transformed set whose dynamics is described by Einstein field equations
is called Einstein conformal frame. The dynamical
equivalence of Jordan and Einstein conformal frames does not
generally imply that they are also physically equivalent.  In fact
it is shown that some physical systems can be differently
interpreted in different conformal frames \cite {soko} \cite{no}.
The physical status of the two conformal frames is an open
question which we are not going to address here.\\
We will work in Einstein conformal frame.  The motivation is that in this frame there is a coupling between the scalar degree of freedom and matter
sector induced by the conformal transformation.  In this context, we have already studied the coincidence problem without any use of holographic principle \cite{bi}.  It is shown that
the requirement of a constant ratio of energy densities of the two components, puts some constraints on the functional form of the $f(R)$ function.  In other terms, there is a class of parameterized $f(R)$ models for which the coincidence problem may be alleviated within a particular region of the parameters space.\\  Applying the holographic principle to $f(R)$ gravity leads to some important changes in this result.  We
 show that identification of IR cutoff with the Hubble scale in a spatially flat universe necessarily leads to a stationary ratio of energy densities corresponding to dark energy and matter sector regardless of the functional form of the $f(R)$ function.  Thus, \emph{any} holographic
$f(R)$ model can address the coincidence problem in a spatially flat universe.
In a universe with a spatial curvature, we will show that the ratio of
energy densities is no longer a constant and evolves during expansion of the universe.  We argue that alleviation of the coincidence problem requires conditions that directly link to the cosmological viability of  $f(R)$ models. \\
Let us start with introducing the action for an $f(R)$
gravity theory in the Jordan frame
\begin{equation}
S_{JF}= \frac{1}{2}\int d^{4}x \sqrt{-g}~M_p^2~ f(R) +S_{m}(g_{\mu\nu}, \psi)\label{b1}\end{equation}
where $g$ is the
determinant of $g_{\mu\nu}$ and $S_{m}$ is the action
of (dark) matter which depends on the metric $g_{\mu\nu}$ and some (dark) matter
field $\psi$.  Stability in matter sector (the Dolgov-Kawasaki
instability \cite{dk}) imposes some conditions on the functional
form of $f(R)$ models.  These
conditions require that the first and the second derivatives of
$f(R)$ function with respect to the Ricci scalar $R$ should be
positive definite.  The positivity of the first derivative ensures
that the scalar degree of freedom is not tachyonic and positivity
of the second derivative tells us that graviton is not a ghost.\\
It is well-known that $f(R)$ models are equivalent to models in which
a scalar field minimally couples to gravity with an appropriate
potential function.  In fact, we may use a new set of variables
\begin{equation}
\bar{g}_{\mu\nu} =\Omega~ g_{\mu\nu} \label{b2}\end{equation}
\begin{equation} \phi = \frac{M_p}{2\beta} \ln \Omega
\label{b3}\end{equation}
 where
$\Omega\equiv\frac{df}{dR}=f^{'}(R)$ and
$\beta=\sqrt{\frac{1}{6}}$. This is indeed a conformal
transformation which transforms the above action in the Jordan
frame to the following action in the Einstein frame \cite{soko} \cite{w}
\begin{equation}
S_{EF}=\frac{1}{2} \int d^{4}x \sqrt{-\bar{g}}~\{\frac{1}{M_p^2}
\bar{R}-\bar{g}^{\mu\nu} \nabla_{\mu} \phi~ \nabla_{\nu} \phi
-2V(\phi)\}+ S_{m}(\bar{g}_{\mu\nu}e^{2\beta \phi/M_p}
, \psi) \label{b4}\end{equation}All indices are raised and lowered by $\bar{g}_{\mu\nu}$.  In the Einstein frame, $\phi$ is a minimally
coupled scalar field with a self-interacting potential which is
given by
\begin{equation}
V(\phi(R))=\frac{M_p^2(Rf'(R)-f(R))}{2f'^2(R)}
\label{b5}\end{equation} Note that the conformal transformation
induces the coupling of the scalar field $\phi$ with the matter
sector. The strength of this coupling $\beta$, is fixed to be
$\sqrt{\frac{1}{6}}$ and is the same for all types of matter
fields.  In the action (\ref{b4}), we take $\bar{g}^{\mu\nu}$ and
$\phi$ as two independent field variables and variations of the
action yield the corresponding dynamical field equations.
Variation with respect to the metric tensor $\bar{g}^{\mu\nu}$,
leads to
\begin{equation}
\bar{G}_{\mu\nu}=M_p^{-2}~(\bar{T}^{\phi}_{\mu\nu}+
\bar{T}^{m}_{\mu\nu}) \label{b6}
\label{b7}\end{equation} where
\begin{equation}
\bar{T}^{\phi}_{\mu\nu}=\nabla_{\mu}\phi
\nabla_{\nu}\phi-\frac{1}{2}\bar{g}_{\mu\nu}\nabla^{\gamma}\phi
\nabla_{\gamma}\phi-V(\phi)\bar{g}_{\mu\nu}
\label{b8}\end{equation}
\begin{equation}
\bar{T}^m_{\mu\nu}=\frac{-2}{\sqrt{-\bar{g}}}\frac{\delta S_{m}(\bar{g}_{\mu\nu}, \psi)}{\delta \bar{g}^{\mu\nu}} \label{b9}\end{equation} are
stress-tensors of the scalar field and the matter field system.
It is important to note that the two
stress-tensors $\bar{T}^m_{\mu\nu}$ and $\bar{T}^{\phi}_{\mu\nu}$
are not separately conserved.
Instead they satisfy the following equations
\begin{equation}
\bar{\nabla}^{\mu}\bar{T}^{m}_{\mu\nu}=-\bar{\nabla}^{\mu}\bar{T}^{\phi}_{\mu\nu}= \frac{\beta}{M_p} \nabla_{\nu}\phi~\bar{T}^{m}\label{b13}\end{equation} We apply the field equations in a
 homogeneous and isotropic cosmology described by
Friedmann-Robertson-Walker spacetime
\begin{equation}
ds^2=-dt^2+a^2(t)\{ \frac{dr^2}{1-kr^2}+r^2(d\theta^2+\sin ^2\theta d\phi^2)\}
\end{equation}
where $a(t)$ is the scale factor and $k$ determines the spatial curvature. The parameter $k$ can take $0,-1,+1$ corresponding
to spatially flat, open and curved universes, respectively.  We take
$\bar{T}^m_{\mu\nu}$ and $\bar{T}^{\phi}_{\mu\nu}$ as the stress-tensors of a pressureless perfect fluid with energy density
$\bar{\rho}_{m}$, and a perfect fluid with energy density
$\rho_{\phi}=\frac{1}{2}\dot{\phi}^2+V(\phi)$ and pressure
$p_{\phi}=\frac{1}{2}\dot{\phi}^2-V(\phi)$.  In this
case, the equations (\ref{b7}) take the form \footnote{Hereafter we will use unbarred characters in the Einstein frame.}
\begin{equation}
3H^2+\frac{3k}{a^2}=M_p^{-2}(\rho_{\phi}+\rho_{m})
\label{b14}\end{equation}
\begin{equation}
2\dot{H}=-M_p^{-2}[(\omega_{\phi}+1)\rho_{\phi}+\rho_m]+\frac{2k}{a^2}
\label{b14-1}\end{equation}
 where
$\omega_{\phi}=\frac{p_{\phi}}{\rho_{\phi}}$ is equation of state parameter of the scalar field $\phi$, and overdot indicates differentiation with respect
to cosmic time $t$.  We may write the field equations in terms of relative densities defined by
$\Omega_m=\frac{\rho_m}{\rho_c}$ and $\Omega_{\phi}=\frac{\rho_{\phi}}{\rho_c}$ where $\rho_c=3M_p^2H^2$ is the critical density.  These
relative densities then satisfy
\begin{equation}
\Omega_m+\Omega_{\phi}-\Omega_k=1
\label{ome}\end{equation}
where $\Omega_k=\frac{k}{a^2H^2}$.
The conservation equations (\ref{b13}) give
\begin{equation}
\dot{\rho}_{m}+3H\rho_{m}=Q \label{b17}\end{equation}
\begin{equation}
\dot{\rho}_{\phi}+3H(\omega_{\phi}+1)\rho_{\phi}=-Q
\label{b18}\end{equation} where
\begin{equation}
Q=\frac{\beta}{M_p} \dot{\phi}\rho_{m}
\label{b-18}\end{equation} is the interaction term.  This term
vanishes only for $\phi$~=~const., which due to (\ref{b3}) it happens
when $f(R)$ linearly depends on $R$. The direction of energy
transfer depends on the sign of $Q$ or $\dot{\phi}$.  For
$\dot{\phi}>0$, the energy transfer is from dark energy to dark
matter and for $\dot{\phi}<0$ the reverse is true.\\
Let us consider time evolution of the ratio $r\equiv \rho_{m}/\rho_{\phi}$ ,
\begin{equation}
\dot{r}=\frac{\dot{\rho}_{m}}{\rho_{\phi}}-r\frac{\dot{\rho}_{\phi}}{\rho_{\phi}}
\label{c1}\end{equation}  If we combine the latter with the
balance equations (\ref{b17}) and (\ref{b18}), we obtain
\begin{equation}
\dot{r}=3Hr
[\omega_{\phi}+(1+\frac{1}{r})\frac{\Gamma}{3H}]
\label{c2}\end{equation} where
\begin{equation}
\Gamma=\frac{Q}{\rho_{\phi}}=\frac{\beta}{M_p}r\dot{\phi}
\label{c2c}\end{equation}
is the decay rate.  Now we apply the holographic relation to dark energy density $\rho_{\phi}$ with $L=H^{-1}$,
\begin{equation}
\rho_{\phi}=3c^2 M_p^2 H^2 \label{c3}\end{equation} This gives
\begin{equation}
\dot{\rho}_{\phi}=6c^2 M_p^2 H \dot{H}
\label{c4}\end{equation}
We combine (\ref{c3}) with (\ref{b14-1}) to obtain
\begin{equation}
\dot{H}=-\frac{3}{2}H^2(1+\frac{\omega_{\phi}}{r+1})(r+1)c^2+\frac{k}{a^2}
\label{c55}\end{equation}
One can easily check that
\begin{equation}
(r+1)c^2=\Omega_k+1
\label{c5c}\end{equation}
which reduces (\ref{c55}) to
\begin{equation}
\dot{H}=-\frac{3}{2}H^2(1+\frac{\omega_{\phi}}{r+1})(\Omega_k+1)+\frac{k}{a^2}
\label{c5}\end{equation}
Substituting this into (\ref{c4}) gives
\begin{equation}
\dot{\rho}_{\phi}=-9c^2 M_p^2 H^3 (1+\frac{\omega_{\phi}}{r+1})(\Omega_k+1)+6c^2M_p^2H\frac{k}{a^2}
\label{c6}\end{equation}
When we put the latter together with the holographic relation (\ref{c3}) into the balance equation (\ref{b18}), we obtain
\begin{equation}
\omega_{\phi}=\frac{(r+1)}{3(\Omega_k-r)}(\frac{\Gamma}{H}-\Omega_k)
\label{c7-1}\end{equation}
Note that there is no non-interacting
limit in our case since $\Gamma=0$ corresponds to $\phi$~=~const., or equivalently, $\Lambda$CDM model. We may use (\ref{c7-1}) in
the relation (\ref{c2}) to obtain
\begin{equation}
\dot{r}=H(r+1)(\frac{\Gamma}{H}-r)\frac{\Omega_k}{\Omega_k-r}
\label{hh}\end{equation}
For a further step, we introduce the
deceleration parameter which is given by
\begin{equation}
q=-1-\frac{\dot{H}}{H^2}
\end{equation}
This together with (\ref{c5}) and (\ref{c7-1}) results in
\begin{equation}
q=\frac{1}{2}(\Omega_k+1)(\frac{\frac{\Gamma}{H}-r}{\Omega_k-r})
\label{q}\end{equation}
From the above expressions, we note that the important quantities $\omega_{\phi}$, $\dot{r}$ and $q$ which describe evolution
of the universe appear as functions
of $\Gamma$, $r$ and $\Omega_k$.
We will consider two different cases :\\
1) In a spatially flat universe, $\Omega_k=0$ and the relation
(\ref{hh}) results in $\dot{r}=0$ or $r=$~constant.  It is important to note that this result is independent
of the details of the decay rate $\Gamma$ or the interaction $Q$.  Since the interaction is given by the shape of the $f(R)$ function
we conclude that applying holographic principle to the dark energy density $\rho_{\phi}=3c^2 M_p^2 H^2$ in a spatially flat universe necessarily leads to a constant
ratio of energy densities $r=\rho_m/\rho_{\phi}$, irrespective of the form of the $f(R)$ function.  The reasoning is simple : from the holographic
relation (\ref{c3}) one infers that $\rho_{\phi}$  scales like the critical density $\rho_c=3M_p^2H^2$.  As a consequence, the relative density corresponding to $\phi$ must be a constant so that $\Omega_{\phi}=\frac{\rho_{\phi}}{\rho_c}=c^2$.  With this result and the fact that
in a spatially flat universe $\Omega_k=0$, the Friedmann equation (\ref{ome}) results in $\Omega_m=1-c^2$. Thus $\rho_m$ has the same scaling
as $\rho_{\phi}$ and the ratio $r$ is a constant.  \\In the flat case, the relations (\ref{c7-1}) and
(\ref{q}) reduce to
\begin{equation}
\omega_{\phi}=-(1+\frac{1}{r})\frac{\Gamma}{3H}
\label{c7-2}\end{equation}
\begin{equation}
q=\frac{1}{2r}(r-\frac{\Gamma}{H})
\label{q1}\end{equation}
Contrary to the constancy of $r$, accelerating expansion requires
particular configurations of the $f(R)$ function.  This is clear from the expression (\ref{q1}) which the requirement that $q<0$ automatically
sets a constraint on the decay rate.\\
The relation (\ref{q1}) indicates a signature flip of the deceleration parameter during expansion of the universe and evolution
of $\frac{\Gamma}{H}$.  When $\frac{\Gamma}{H}<r$, we have $q>0$ and the expansion is decelerating.  The condition $\frac{\Gamma}{H}=r$ defines
a transition
 region in which $q=0$ and the universe enters the accelerating phase for which $\frac{\Gamma}{H}>r$. It should be noted that the deceleration parameter corresponding to
a viable $f(R)$ gravity model should exhibit this signature flip in a special way so that
\begin{equation}
\frac{dq(z)}{dz}>0~~~~~ for~~~~~ \frac{\Gamma}{H}=r
\label{gam}\end{equation}  It
means that we should have an accelerating phase after the decelerating one in the expansion history of the universe.  Moreover, the transition region
$\frac{\Gamma}{H}=r$ should lie in recent past.  These conditions put some constraints on the shape of $f(R)$ functions.  This situation
is very similar to the case presented in \cite{ref} when the Hubble scale is taken as the cutoff length.  In
that case accelerated expansion constrains the shape of the interaction term of an interacting holographic dark energy model.\\
2) Although a flat universe is usually assumed, it is still quite possible that there is a contribution to the field equations from the spatial curvature \cite{kk}. In a spatially curved universe, the relation (\ref{q}) indicates that $\frac{\Gamma}{H}=r$ remains as the point in which $q$ changes its sign.  \\ If we combine (\ref{hh}) and (\ref{q}) and then use the relation (\ref{ome}), we arrive at
\begin{equation}
\dot{r}=2q(r+1)H\frac{\Omega_k}{\Omega_k+1}=2qH\frac{\Omega_k}{\Omega_{\phi}}
\label{rq}\end{equation}
which relates evolution of $r$ to the deceleration parameter $q$ and $\Omega_k$.  This implies that $\dot{r}$ changes its sign at the same
point that $q$ does, namely at $\frac{\Gamma}{H}=r$.  As the relations (\ref{hh}) and (\ref{rq}) imply, contrary to the spatially flat case, $r$ is not a constant and evolves during expansion of the universe.  Although the density
parameter $\Omega_{\phi}$ is still a constant
equal to $c^2$, due to the non-vanishing spatial curvature the Friedmann equation (\ref{ome}) generally gives different scalings for $\rho_{m}$ and
$\rho_{\phi}$.  The rate of change of the ratio $r$ is closely related to the sign of $q$.  In a spatially closed universe ($\Omega_k>0$) the ratio $r$ is an increasing
function of time when the expansion is decelerating ($q>0$) and it is a decreasing function when the expansion is accelerating ($q<0$).  In a spatially
open universe ($\Omega_k<0$) the reverse is true.\\ There are two important contributions to the changes of $r$, namely, from $\Omega_k$ and
$q$.  There are some indications \cite{kk} that contributions from the spatial curvature should be very small ($|\Omega_k| \sim 0.001$).  On the other hand, changes of $r$ do not also receive strong contributions from $q$ by noting the fact that accelerating expansion
of the universe is a recent phenomenon.  To clarify this point we first note that transition to the accelerating phase takes place when $\frac{\Gamma}{H}=r$.  Assuming that this condition is satisfied in an appropriately small redshift near the present epoch we conclude that
 we live in an epoch in which $\frac{\Gamma}{H}\approx r$.  The relation (\ref{hh}) then indicates that $\dot{r}$ receives a small contribution from the deceleration parameter.  This is consistent with the alleviation
 of the coincidence problem since besides the possibility that the present epoch may be a stationary regime at which the ratio $r$ is a constant, it is also quite possible that we live in a very special epoch, a transient epoch at which the ratio varies slowly with respect to the expansion of the universe \footnote{The relation (\ref{rq}) is equivalent to $\dot{r}=\frac{2q\Omega_k}{c^2}H$.  Following the above discussion we may write
 $\frac{2q\Omega_k}{c^2}<<1$ and then $\dot{r}<< H$, the rate of change of $r$ is slower than expansion of the universe.}.\\
 In summary, applying holographic principle to $f(R)$ models in a spatially flat universe necessarily leads to a stationary
 energy density ratio independent of the form of the $f(R)$ function.  In a spatially curved universe, although $r$ does not appear as a constant ratio and details of its evolution is model-dependent, \emph{all} cosmologically viable $f(R)$ models for which
 the transition region $\frac{\Gamma}{H}=r$ lies in recent past lead to
 $r\approx$~constant and can alleviate the coincidence problem.


\vspace{3cm}
\begin{thebibliography}{99}

\bibitem{win} S. Weinberg, Rev. Mod. Phys. {\bf 61}, 1 (1989)
\bibitem{R} A. G. Riess, Astrophys. J. {\bf 560}, 49, (2001)
\bibitem{b}Y. Bisabr and H. Salehi, Class. Quantum Grav. {\bf 19}, 2369 (2002)\\
Y. Bisabr, Gen. Relativ. Gravit. {\bf 42}, 1211 (2010)
\bibitem{r}S. M. Carroll, A. De Felice, V. Duvvuri, D. A. Easson, M.
Trodden
and M. S. Turner, Phys. Rev. D {\bf 71}, 063513, (2005)\\
G. Allemandi, A. Browiec and M. Francaviglia, Phys. Rev. D {\bf 70}, 103503 (2004)\\
X. Meng and P. Wang, Class. Quantum Grav. {\bf 21}, 951 (2004)\\
M. E. soussa and R. P. Woodard, Gen. Rel. Grav. {\bf 36}, 855
(2004)\\
S. Nojiri and S. D. Odintsov, Phys. Rev. D {\bf 68}, 123512, (2003)
\bibitem{suss} G. 't Hooft, gr-qc/9310026\\
L. Susskind, J. Math. Phys. {\bf 36}, 6377, (1995)
\bibitem{bek} J. D. Bekenstein, Phys. Rev. D {\bf 7}, 2333, (1973)\\
 J. D. Bekenstein, Phys. Rev. D {\bf 9}, 3292, (1974)\\
 J. D. Bekenstein, Phys. Rev. D {\bf 23}, 287, (1981)\\
 J. D. Bekenstein, Phys. Rev. D {\bf 49}, 1912, (1994)
\bibitem{co}A. G. Cohen, D. B. Kaplan and A. E. Nelson, Phys. Rev.
Letts. {\bf 82}, 4971, (1999)
\bibitem{li}H. D. S. Hsu, Phys. Letts. B {\bf 594}, 13, (2004)\\
M. Li, Phys. Letts. B {\bf 603}, 1, (2004)
\bibitem{pp}D. Pavon and W. Zimdahl, Phys. Letts. B {\bf 628}, 206
(2005)\\
D. Pavon and W. Zimdahl, Class. Quantum Grav. {\bf 24}, 5461 (2007)
\bibitem{ref}S. del Campo, J. o. C. Fabris, R. Herrera and W. Zimdahl, Phys. Rev. D {\bf 83}, 123006 (2011)
\bibitem{soko} G. Magnano and L. M. Sokolowski, Phys. Rev. D {\bf 50}, 5039 (1994)
\bibitem{no}Y. M. Cho, Class. Quantum Grav. {\bf 14}, 2963
(1997)\\
E. Elizalde, S. Nojiri and S. D. Odintsov, Phys. Rev. D {\bf 70},
043539 (2004)\\
 S. Nojiri and S. D. Odintsov, Phys. Rev. D {\bf 74},
086005
(2006)\\
S. Capozziello, S. Nojiri, S. D. Odintsov and A. Troisi, Phys.
Lett. B {\bf 639}, 135 (2006)\\
\bibitem{bi}Y. Bisabr, Phys. Rev. D {\bf 82}, 124041 (2010)
\bibitem{dk}A. D. Dolgov, and M. Kawasaki, Phys. Lett. B {\bf
573},1 (2003)
\bibitem{w}K. Maeda, Phys. Rev. D {\bf 39}, 3159 (1989)\\
D. Wands, Class. Quant. Grav. {\bf 11}, 269 (1994)
\bibitem{kk}M. Tegmark et al, Phys. Rev. D {\bf 69}, 103501 (2004)\\
K. Ichikawa, M. Kawasaki, T. Sekiguchi and T. Takahashi, JCAP 0612, 005 (2006)\\
U. Seljak, A. Slosar and P. Mc Donald, JCAP 0610, 014 (2006)\\
Y. Wang and P. Mukherjee, Phys. Rev. D {\bf 76}, 103533 (2007)



\end{thebibliography}
\end{document}